\documentstyle[aps,preprint]{revtex}

\tighten

\begin{document} \draft \title{New Discrete Basis for Nuclear Structure
Studies}
\author{M.V.~Stoitsov,$^{1,2}$ W. Nazarewicz,$^{3-5}$ and  S.~Pittel,$^{6}$}

\address{$^1$Institute of Nuclear Research and Nuclear Energy, Bulgarian
Academy
of Sciences
 \\Sofia-1784, Bulgaria}
\address{$^2$Joint Institute for Heavy Ion
Research, Oak Ridge, Tennessee 37831}
\address{$^3$Department of Physics,
University of Tennessee, Knoxville, Tennessee 37996}
\address{$^4$Physics
Division, Oak Ridge National Laboratory, Oak Ridge, Tennessee 37831}
\address{$^5$Institute of Theoretical Physics, University of Warsaw, ul.
Ho\.za
69, PL-00-681 Warsaw, Poland}
\address{$^6$Bartol Research Institute, University
of Delaware, Newark, Delaware 19716}

\date{\today}

\maketitle

\begin{abstract}
A complete discrete set of  spherical single-particle wave functions for
studies
of  weakly-bound many-body systems is proposed. The new basis is obtained by
means of a local-scale point transformation of the spherical harmonic
oscillator
wave functions. Unlike the harmonic oscillator states, the new wave
functions decay exponentially at large distances. Using the new basis,
characteristics of weakly-bound orbitals are analyzed and the ground state
properties of some spherical doubly-magic  nuclei are studied. The basis of
the
transformed harmonic oscillator is a significant improvement over the
harmonic
oscillator basis, especially in studies of exotic nuclei where the coupling
to
the particle continuum is important. \end{abstract}

\pacs{PACS number(s): 21.10.Pc, 21.60.Cs, 21.60.Jz, 71.15.Mb}

\narrowtext

\section{Introduction}

Understanding the nature of exotic nuclei with extreme isospin values is one
of
the most exciting challenges of current nuclear structure physics. Thanks to
developments in radioactive ion beam  instrumentation, we are in the process
of
exploring the very limits of nuclear binding, namely the regions of the
periodic
chart in the neighborhood of the particle drip lines. Several new structural
features arise in the description of nuclei near the drip lines.
Firstly,
such nuclei exhibit very weak binding, which leads to extended spatial
distributions.
Secondly, the particle continuum plays a critical role in a
description of the properties of these nuclei. Unlike more stable nuclei
closer
to the valley of beta stability, a proper theoretical description of
weakly-bound
systems (such as halo nuclei) requires a very careful treatment of the
asymptotic
part of the nucleonic density.

The correct treatment of the particle continuum and near-threshold (e.g.
halo) states poses  a significant
theoretical challenge. One possible way of tackling this problem is by means
of
the continuum shell model (CSM), in which the basis consists of both
bound and unbound states, e.g., the eigenvectors of some
finite shell-model potential
\cite{[Fan61],[Glo67],[Phi77a],[Mic78],[Hal80],[Isk91],[Ben98]}. An
alternative
approach
is to discretize the continuum by means of
Sturmian function expansions or resonant state expansions. Sturmian
functions  form a discrete set of states which
behave asymptotically as outgoing waves; they have been used as a basis in
the
solution of scattering equations, including various applications of the CSM
\cite{[Glo67],[Vaa79],[Raw82]}. The Gamow
(Berggren) states are the  eigenstates of the time-independent
Schr\"odinger equation
with complex eigenvalues \cite{[Rom72],[Ber68],[Ver87]}; they have been
applied
to many problems involving an
unbound spectrum \cite{[Lin94],[Ber96a],[For97]}.
Another possibility  is to employ
the canonical Hartree-Fock-Bogolyubov (HFB) basis of the
independent-quasiparticle Hamiltonian \cite{[Dob84],[Dob96]}. The
canonical states, i.e., the eigenstates of the
one-body density matrix, form a
complete localized basis with proper one-body asymptotics.
Unfortunately, in order to find
the canonical basis, the self-consistent one-body
density matrix must be known,
and this is not always possible. Finally, the particle continuum can be
discretized by  placing the nucleus inside a very large
box. Since the properties of
the nucleus itself must not depend
on the  box size, one has to subtract the
contribution from the free-gas states that are introduced
\cite{[Tub79],[Ker81a],[Dea85]}. The coordinate-space HFB
\cite{[Dob84],[Dob96],[Ter96],[Rei97]} and relativistic
Hartree-Bogolyubov (RHB) \cite{[Poe97]} methods are based on this principle.

One of the most important tools of nuclear structure physics is the harmonic
oscillator (HO) potential. The simple analytic structure of HO wave functions
greatly simplifies shell-model studies \cite{[Hey90],[osc91],[Tal93]}. Since
the
HO wave functions form a complete {\em discrete} set, they typically serve
as the
single-particle basis of choice in microscopic many-body calculations.
Another useful feature of HO states is that they are all spatially localized,
so that the resulting densities and currents are localized as well. The
disadvantage of HO wave functions is that, because of their gaussian
asymptotics,
they cannot describe the falloff of nuclear density distributions near the
nuclear surface and beyond. To get the falloff correct, it is essential to
mix HO
basis states. This is a serious problem when dealing with weakly-bound
nuclei,
where the density distributions fall off very slowly and the bases required
are
extremely large and thus very difficult  to handle.  Hence, the HO
basis is not particularly useful when dealing with weakly-bound nuclei.

It is desirable, therefore, to have an alternative to HO wave functions for
use
in nuclear structure studies. Ideally, the new basis should preserve as
many of the advantageous features of HO wave functions as possible,  but, at
the
same time, it should have an improved asymptotic behavior. In this work, we
consider a new spherical single-particle basis, obtained through the
application
of a general local-scaling point transformation to harmonic oscillator wave
functions. This approach was originally developed by Petkov and Stoitsov,
and is
described in detail in Ref.~\cite{[Pet83a]}. In this study,  we pay
particular
attention to the asymptotic form of the resulting transformed  HO wave
functions,
developing the method so as to guarantee an appropriate exponential falloff.
This
makes the new basis especially useful in the description of weakly-bound
nuclear
systems.

The paper is organized as follows. Section \ref{basis} briefly reviews the
method
of Petkov and Stoitsov, and introduces the transformed HO
basis. Stability
tests of the new basis, when applied to weakly-bound orbitals, are contained
in
Sec.~\ref{tests}. In Sec.~\ref{HF}, we use the new basis in  variational
calculations based on the density functional theory. Finally,
Sec.~\ref{conclusions} summarizes the principal conclusions of the work and
spells out some issues for future consideration.

\section{Transformed Oscillator Basis}\label{basis}

\subsection{Local-scale point transformations}

The key ingredient in the construction of our new basis is a coordinate
transformation based on the local-scale transformation method
\cite{[Pet83a]}. A
local-scaling point transformation (LST) replaces the original coordinate
${\bf
r}$ by a new coordinate ${\bf r}^{\prime }={\bf f}({\bf r})\equiv \hat{\bf
r}f({\bf r)}$. The new coordinate is in the same direction as ${\bf r}$, but
has
a new magnitude $r^{\prime }=f({\bf r)}$, depending on a scalar function
$f({\bf
r)}$ (called the LST function). It is assumed that $f({\bf r)}$ is an
increasing
function of $r$ and $f(\bf{0})$=0. The set of invertible transformations of
this
type forms a LST group.

Given a model $A$-particle wave function $\bar{\Psi}({\bf r}_{1},{\bf
r}_{2},...,{\bf r}_{A})$,  the LST
transforms it into a new wave function

\begin{equation}\label{lstwf}
\Psi_{f}({\bf r}_{1},{\bf r}_{2},...,{\bf r}_{A})=\left[ \prod_{i=1}^{A}
\frac{f^2({\bf r}_i)}{r^2_i}\frac{\partial f({\bf r}_i)}{\partial r_i}
\right]^{1/2}\bar{\Psi}({\bf f}({\bf r}_{1}),{\bf f}({\bf r}_{2}),...,{\bf f}
({\bf r}_{A})).
\end{equation}
Assuming that the model wave function is normalized to unity,
\begin{equation}
\langle \bar{\Psi}|\bar{\Psi} \rangle =1,
\end{equation}
the LST wave function $\Psi_{f}({\bf r}_{1},{\bf r}_{2},...,{\bf r}_{A})$
will
also be normalized to unity, regardless of the choice of $f({\bf r})$.

The local one-body density corresponding to an $A$-body wave function
${\Psi}$
is
\begin{equation}
 {\rho}({\bf r})=A\int |\Psi
({\bf r},{\bf r}_{2},...,{\bf r}_{A})|^{2}
d{\bf r}_{2},...,d{\bf r}_{A} ~.
\label{modden}
\end{equation}
It then follows from (\ref{lstwf}) that there
exists a simple relation between the local density $\rho_f(\bf r)$ associated
with the LST function ${\Psi}_f$ and the model local density $\bar{\rho}(\bf
r)$
corresponding to the model function  $\bar{\Psi}$:
\begin{equation}
\label{lstden}
\rho_f({\bf r})=\frac{f^{2}({\bf r})}{r^{2}}\frac{\partial f({\bf
r})} {\partial r}\bar{\rho}({\bf f}({\bf r)}) ~.
\end{equation}
The relation
(\ref{lstden}) is particularly useful when the density $\rho_f(\bf r)$ is
known (or at least  approximately known). Given a model wave
function, Eq.~(\ref{lstden}) becomes a first-order nonlinear differential
equation for the LST function $f$. For a spherically symmetric system,
$\rho_f$, $\bar{\rho}$, and $f$ depend only on  $r$=$|{\bf r|}$, and
Eq.~(\ref{lstden}) can be reduced to the nonlinear algebraic equation
\begin{equation}\label{lsteq}
\int\limits_{0}^{r}\rho_f
(u)u^{2}du=\int\limits_{0}^{f(r)}\bar{\rho}(u)u^{2}du ~,
\end{equation}
which can be solved subject to the boundary condition $f(0)=0$.
Such an approach was pursued in a series of works
\cite{[Dim86],[Sto88a],[Dim88]} based on the energy density functional
method.
A review of the density functional theory based on the LST can be found in
the
monograph \cite{[Pet91]}.

In the context of shell-model or mean-field applications, the particularly
interesting case is  when the model function is a  Slater determinant,
\begin{equation}\label{ssmod}
\bar{\Psi}({\bf r}_{1},{\bf r}_{2},...,{\bf r}_{A})=
\frac{1}{\sqrt{A!}}\det |\bar{\psi}_{i}({\bf r}_{j})|  ~,
\end{equation}
built from a complete set of model single-particle wave functions
$\bar{\psi}_{i}({\bf r})$. Due to the unitarity of
transformation (\ref{lstwf}), the LST wave function retains the structure of
a
Slater determinant,
\begin{equation}\label{ssf}
\Psi_{f}({\bf r}_{1},{\bf
r}_{2},...,{\bf r}_{A})=\frac{1} {\sqrt{A!}}\det |\psi_{i}({\bf r}_{j})| ~,
\end{equation}
but with new single-particle wave functions
\begin{equation}
\label{lstspwf} \psi _{i}({\bf r})=\left[\frac{f^{2}({\bf r})}{r^{2}}
\frac{\partial f({\bf r})} {\partial r}\right]^{1/2}\bar{\psi}_{i}({\bf
f}({\bf
r)}) ~.
\end{equation}
These functions, in the following referred to as the LST basis, form a
complete
set of single-particle states.

\subsection{Transformed harmonic oscillator basis}

Having in mind the numerous advantages of the HO basis, we choose for the
model
single-particle wave functions the eigenfunctions of a HO potential. Since in
this work we consider spherically symmetric systems only, the angular part
of the
single-particle wave function is not affected by the LST transformation. The
radial HO wave functions $R^{\rm HO}_{nl}(r)$ are characterized by one
external
parameter, namely the oscillator length $a_{\rm osc}=\sqrt{\hbar/M\omega}$.
The
LST basis associated with HO model wave functions will be referred to
throughout the remainder of this
paper
as the transformed harmonic
oscillator (THO) basis. The states in this basis are given by
Eq.~(\ref{lstspwf}):
\begin{equation} \label{psr} R^{\rm THO}_{nl}(r)=
\left[\frac{f^{2}(r)}{r^{2}}\frac{df(r)}{dr}\right]^{1/2}R^{\rm
HO}_{nl}(f(r)).
\end{equation}

Up to this point, the LST function has not been defined. Ideally, we would
like to parametrize $f(r)$ in
as simple manner as possible consistent with the requirement that it
reproduces
the generic features of the local density $\rho_f$ at small and large
distances
$r$. Of course, the actual behavior of $\rho_f$ is affected by nuclear
shell effects. Therefore, we concentrate on the average behavior of
$\rho_f(r)$
only.

In the nuclear interior, the  average local density
varies rather weakly with $r$. Due to the effects of Coulomb repulsion, the
proton
charge density is expected to exhibit a central depression
\cite{[Fri86]}. Consequently, at low values of $r$, we assume the following
ansatz:
 \begin{equation}\label{lowr} \rho(r) \approx \rho_0+cr^2.
\end{equation}
Eq.~(\ref{lsteq}) can then be cast into a fifth-order polynomial equation
for the
LST function $f(r)$ at small $r$.  To further simplify the problem,  we
assume
that the model density $\tilde\rho(r)$ is constant in the inner region. In
this
limit, the LST function assumes the simple form
\begin{equation}\label{lowf} f(r)
= r (a+br^2)^{1/3} ~,
\end{equation}
where $a$ and $b$ are  parameters still to be determined.

For large values of $r$, i.e., outside the nucleus, the average
LST density should decay exponentially. Hence,
\begin{equation}\label{highr}
\rho(r) \approx
\rho({R})\exp\left(\frac{{R}-r}{\tilde{a}}\right) ~,
\end{equation}
where $R$ is a characteristic LST radius that is
significantly greater than the nuclear radius and $\tilde{a}$ is the LST
diffuseness parameter. In contrast, the model density based on HO states
exhibits
a gaussian asymptotic behavior. Using Eq.~(\ref{lsteq}), we obtain an
approximate expression for $f(r)$ at large distances,

\begin{equation}\label{highf}
f(r) = \sqrt{\frac{d_{-2}}{r^2}+\frac{d_{-1}}{r} +d_0 + d\ln{r} +
d_1r} ~,
\end{equation}
where  $d_{-2},~d_{-1},~d_0,~d_1$, and  $d$  are parameters.
Asymptotically, the linear term in the expansion (\ref{highf}) takes over and
$f(r\rightarrow \infty )\sim r^{1/2}$, as it should.

The parameters
$\{d_{-2},d_{-1},d_{1},d_{0},d\}$ can be determined by the requirement
that $f(r)$ and its first, second,
third, and fourth derivatives are continuous at the
point  $R$.  We are then left with only three independent parameters
$\{a,b,R\}$
in the LST function $f(r)$, which can be chosen to optimize the THO basis
for the
physics problem of interest. In practical calculations, we assume that
$a,b\ge
0$, thereby ensuring that $f(r)$ is a monotonically increasing function of
$r$.

The ansatz (\ref{lowf})-(\ref{highf}) for the LST function guarantees that
all
the THO states  (\ref{psr}) are spatially localized and decay  exponentially
at
large distances. The THO functions are continuous up to their fourth
derivatives.
The HO length is absorbed into the coefficients $a$ and $b$, and does not
appear as an additional parameter. It should be stressed here that the
parametrization of $f(r)$ assumed in our work reflects our desire to keep it
as
simple and practical as possible. Although, as will be demonstrated in the
following sections, this three-parameter form of the LST function performs
very well in actual calculations, other choices are possible and could prove
useful (or even essential) in other applications.

\section{Application of the THO basis to weakly-bound states}\label{tests}

On the neutron-rich side of the valley of stability, there
appear loosely-bound
few-body systems called neutron halo nuclei
(see Refs.~\cite{[Mue93],[Rii94],[Han95],[Tan96]}
for  reviews).
In these nuclei,
weak neutron binding implies large spatial dimensions and
the existence of the halo
(i.e., a dramatic excess of neutrons at large distances).
Theoretically, the weak binding and the corresponding
proximity to the particle continuum, together with
the need for explicit treatment of few-body dynamics,
makes the subject of halos both extremely interesting and difficult.

In this section, we apply the THO basis to weakly-bound
single-particle states to assess its potential usefulness in the description
of
nuclei far from stability and, in particular, halo nuclei. We focus on
spherical
single-particle states and assume that they come from a finite square well
(SQW)
potential with radius $R_0$ and depth $-V_0$. We concentrate on $\ell$=0
orbitals, since they are the best candidates for halos
\cite{[Rii92],[Mis97]}.

More specifically, we carry out a diagonalization of the SQW Hamiltonian
within
a truncated THO single-particle basis and compare the results that emerge
with those of the exact SQW solutions. For the sake of comparison, we also
carry out calculations in the analogous truncated HO basis. All calculations
are
carried out for a fixed radius $R_0$=7.11\,fm of the SQW potential and under
the
assumption that the most weakly-bound $\ell$=0 state is the $3s$. To simulate
scenarios with varying degrees of binding of the $3s$ orbital, we vary the
well
depth. The parameters defining the basis, i.e., the oscillator
length in the HO variant and the $\{a, b, R\}$ parameters of the THO basis,
have been chosen so as to minimize the single-particle energy of the $3s$ halo
state. For a given calculation in either the HO or THO basis, truncation is
defined to include {\em all} single-particle states belonging to $N$$\leq$$N_{\rm
max}$ oscillator shells.

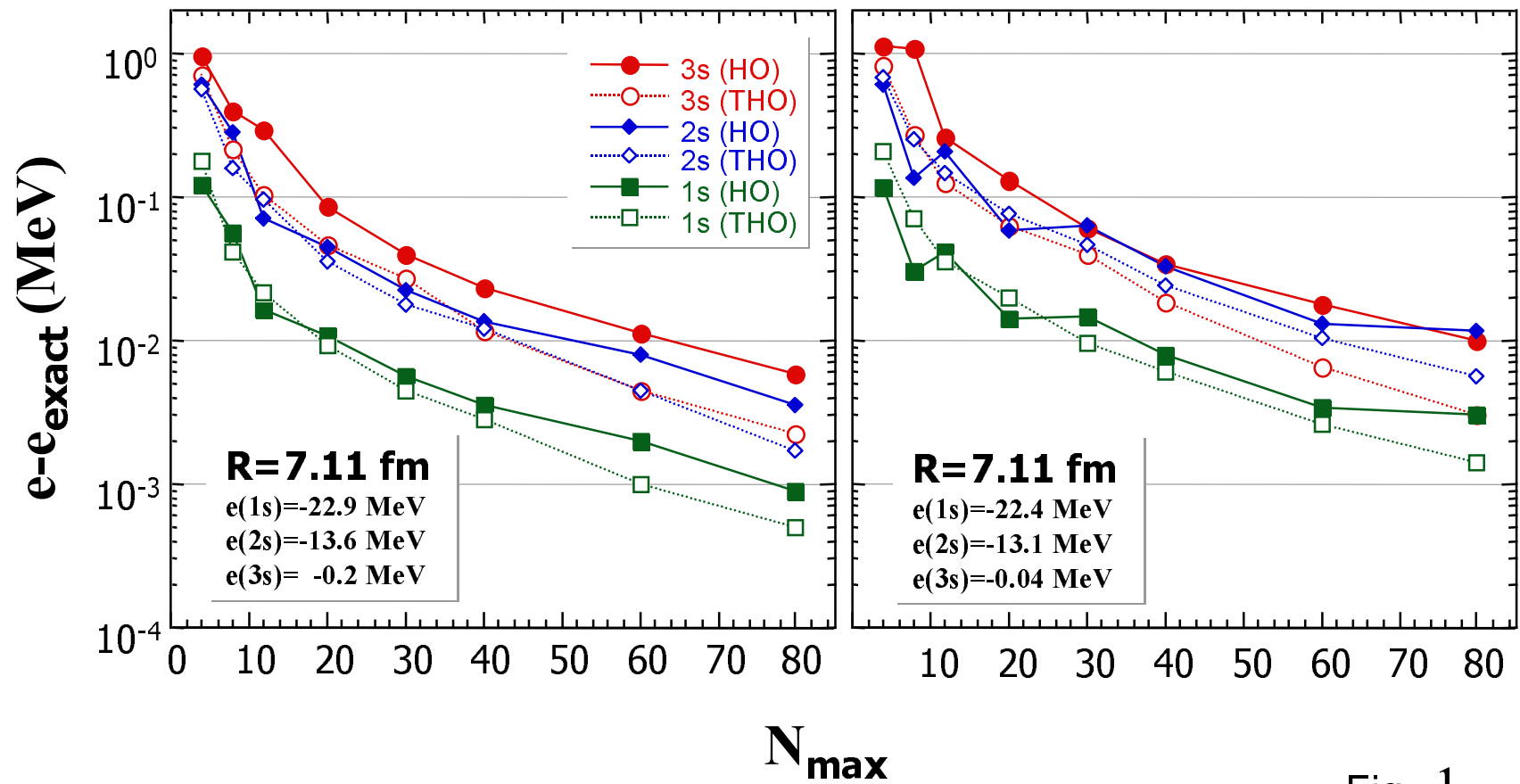
\begin{figure}
\caption{Deviation
between approximate energies based on the basis expansion method and exact
energies of the $1s$, $2s$, and $3s$ states of the SQW Hamiltonian as a
function
of the number of HO quanta included in the basis, $N_{\rm max}$. The solid
lines are the results obtained when expanding in a HO basis; the dotted
lines are the results obtained using the THO basis. The radius of the well
was
taken to be $R_0$=7.11\,fm. The different binding energies of the $3s$ halo
orbital, $-$200 keV (a) and $-$40\,keV (b), were achieved by changing the
well
depth.} \label{esp} \end{figure}

The ability of the HO and THO  basis expansions to reproduce single-particle
energies is illustrated in Fig.~\ref{esp}, which shows the deviation between
approximate and exact energies of the $1s$, $2s$, and $3s$ states as a
function
of $N_{\rm max}$. The energy of the $3s$ state was assumed in these
calculations
to be very low: $-$200 keV (Fig.~\ref{esp}a) and $-$40\,keV
(Fig.~\ref{esp}b). As
can be seen from the figure, the THO basis offers a systematic improvement
over
the traditional HO expansion. As an example, $N_{\rm max}$=20 THO shells are
sufficient to reproduce the  energy of the $3s$ halo state with an  accuracy
of
50-60 keV, whereas $N_{\rm max}$$\approx$30 is required with the HO basis.
For
the well-bound $1s$ and  $2s$ states, there is also an improvement when
using the
THO basis. (The low-$N_{\rm max}$ fluctuations seen in the $1s$ and  $2s$
curves
reflect the fact that the basis was optimized to the energy of the $3s$ halo
state, so that the resulting radial asymptotics is not appropriate for
the more deeply-bound orbitals.)

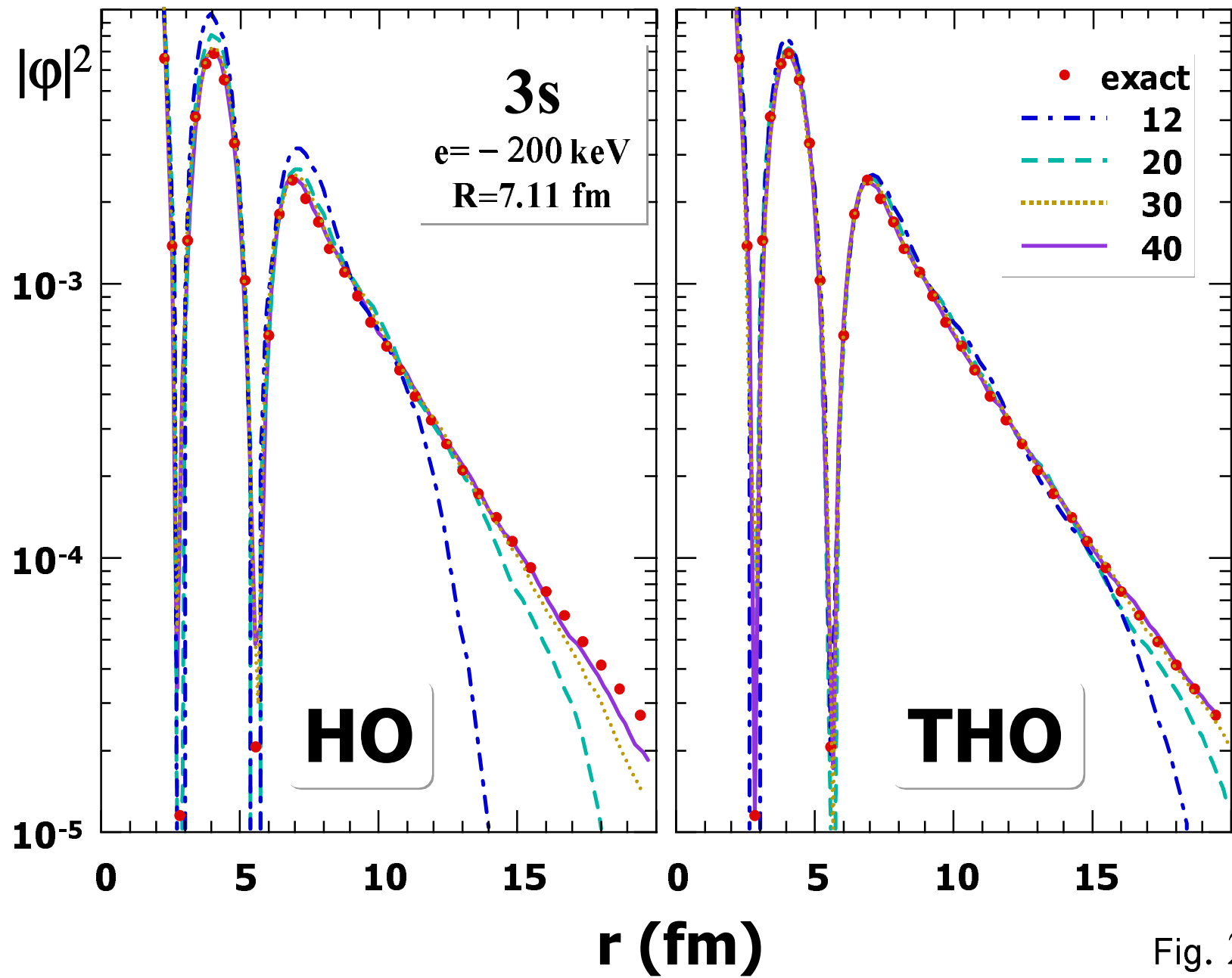
\begin{figure}
\caption{Exact (dots) and approximate (lines) wave functions of the $3s$ halo
state at $e$=$-$200\,keV as a function of $r$. The approximate wave functions
were obtained by means of the basis expansion method
(HO, left panel; THO, right panel) in  $N_{\rm max}$=12,
20, 30, and 40 HO (THO) shells.}
\label{wf1}
\end{figure}

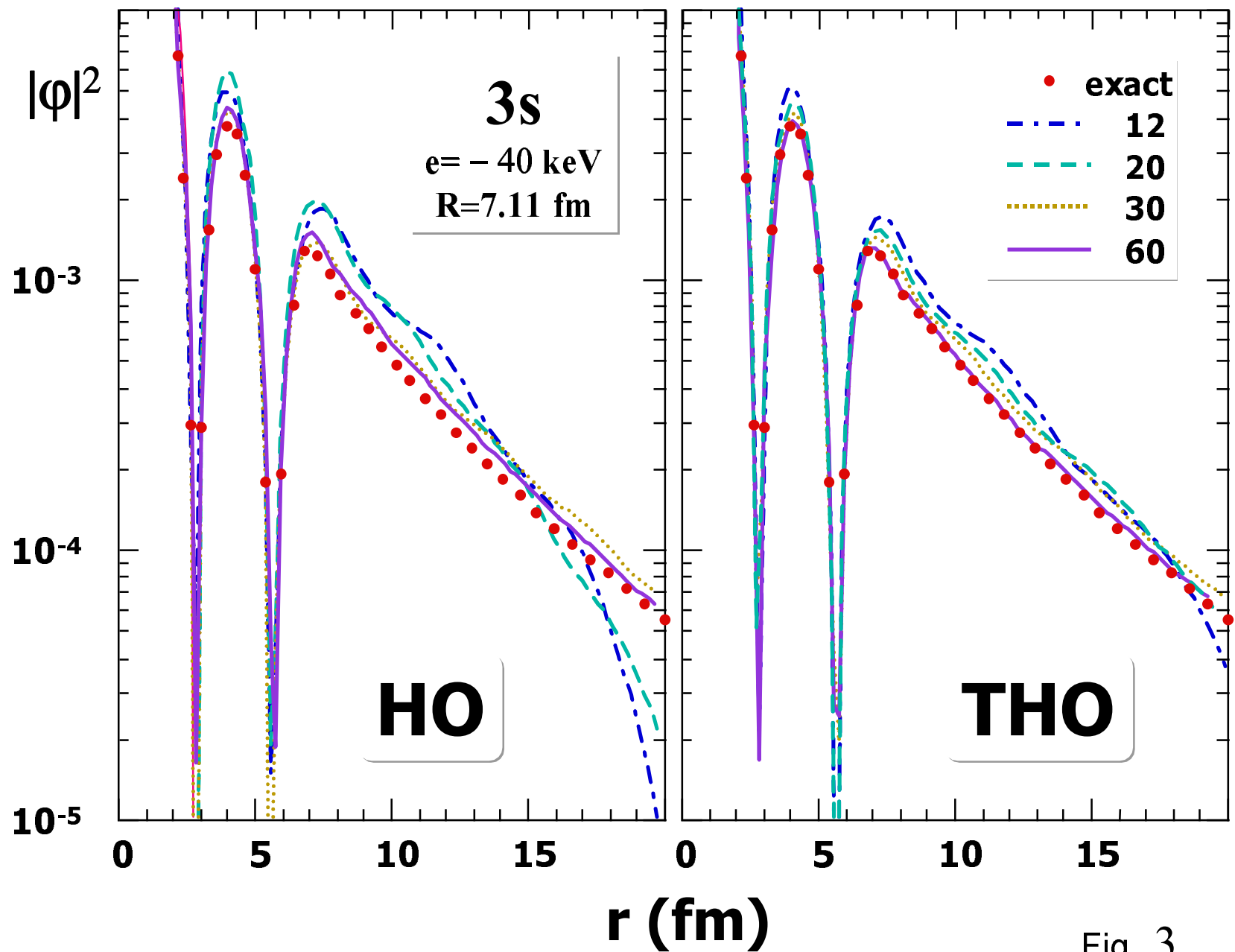
\begin{figure}
\caption{Same as in Fig.~\protect\ref{wf1} but
for the  wave functions of the $3s$ halo
state at $e$=$-$40\,keV and for  $N_{\rm max}$=12,
20, 30, and 60 HO (THO) shells.}
\label{wf2}
\end{figure}

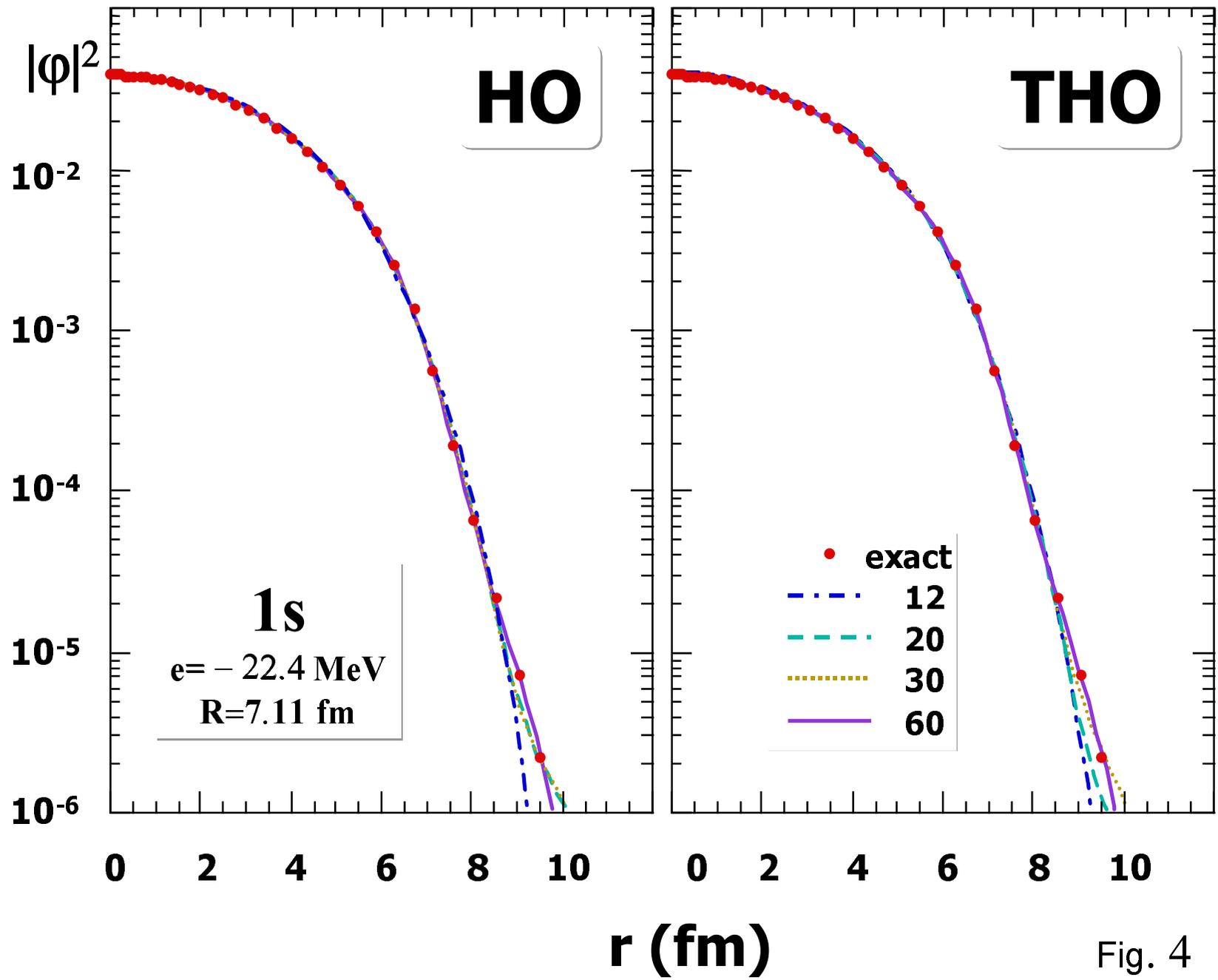
\begin{figure}
\caption{Same as in Fig.~\protect\ref{wf1} but
for the  wave functions of the well-bound $1s$
state at $e$=$-$22.4\,MeV.}
\label{wf3}
\end{figure}

Figures \ref{wf1}-\ref{wf3} compare the exact and approximate wave
functions. It
is gratifying to see that even with a relatively low number of THO shells the
structure of the $3s$ orbital (Figs. \ref{wf1} and \ref{wf2}) is well
reproduced
out to 15-18\,fm, and the agreement with the exact eigenvector becomes
excellent
for $N_{\rm max}$=30. Again, as for the single-particle energies, one
needs at least 30 HO shells to obtain results of comparable quality out to
15-18\,fm. For well-bound states, such as the $1s$ state at --22.4\,MeV shown in
Fig. \ref{wf3}, both expansions work equally well: 20 HO or THO shells are
sufficient to reproduce the exact result in the physically interesting region.

It is instructive to discuss why so many HO shells are required to reproduce
a
wave function out to a very large distance.  In the HO approximation, the
classical radius of an orbit with principal quantum number $N$
and orbital quantum number $\ell$=0 is given by \cite{[Boh69]}
\begin{equation}\label{orbit}
r_{cl} \approx a_{\rm osc} \sqrt{2N}  ~.
\end{equation}
{}From this we see that the radial information contained in HO wave functions
varies quite slowly with the principal quantum number. To build up the
large-$r$
dependence in a radial wave function, we must include HO states of
very high principal quantum numbers. In contrast, the THO basis has no such
restriction on the radial content of its wave functions and convergence can
be
achieved much more rapidly.

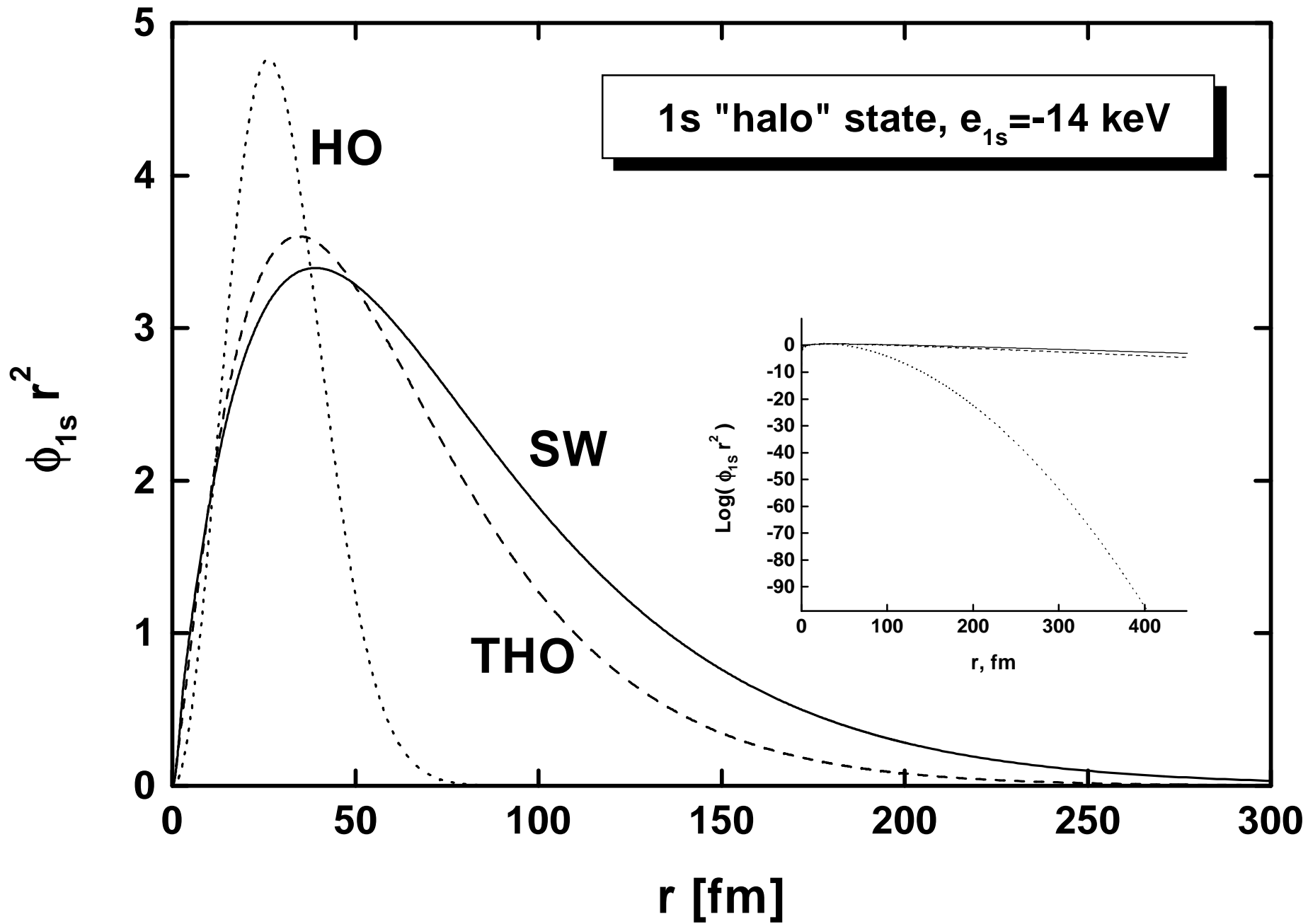
\begin{figure}
\caption{The wave function  of a $1s$ SQW eigenstate
at $e$=--14\,keV (SQW, solid line).
The HO (dotted line) and THO (dashed line) $1s$
wave functions  were determined by maximizing their overlap
with the SQW state. The asymptotic behavior (in logarithmic
scale) is shown in the insert.
}
\label{Na}
\end{figure}

Another rather extreme example is shown in Fig.~\ref{Na}.
Here, we consider the single-particle wave function of a $1s$ state
at $e$=$-$14\,keV.
The SQW wave function has a very large spatial
extension. In this case,
instead of  performing the full basis expansion,
the {\em optimized} single-particle  $1s$ HO and THO
wave functions were obtained by maximizing their respective overlaps
with the exact solution. (In this way one is testing the ability of
unperturbed basis states to reproduce the exact wave functions of
weakly-bound
states.) The corresponding wave functions are plotted in Fig.~\ref{Na}. The
THO
wave function has a squared overlap with the exact SQW solution of $|\langle
{\rm
SQW}|{\rm THO}\rangle|^2$=0.994, whereas the HO wave function has a much
lower
squared overlap of $|\langle {\rm SQW}|{\rm HO}\rangle|^2$=0.877. Clearly,
the THO
basis with the LST functions (\ref{lowf})-(\ref{highf}) is much better able
to
reproduce the tail of a halo wave function than the HO basis.

\section{Variational calculations with Skyrme forces}\label{HF}

In this section, we test whether Slater determinants built up in terms of the
single-particle HO and THO wave functions are able to reproduce the results
of
full self-consistent HF calculations. We follow the energy density
functional approach \cite{[Dim86],[Pet91],[Bra85]}, whereby the total HF
energy,
taken as an expectation value of the nuclear Hamiltonian over a
trial
Slater determinant, involves a sum of the Skyrme  and Coulomb energies. As
usual,
the Skyrme energy density is expressed in terms of local nucleon densities,
kinetic energy densities, and spin-orbit densities, all defined in terms of
the
variational single-particle states. The Coulomb energy density, which
depends on
the local proton density, contains both direct and exchange terms, the latter
taken in the Slater approximation. The effective interaction used in these
calculations was the Skyrme force SkP \cite{[Dob84]}.

In the HO variational analysis, the energy functional of the SkP Hamiltonian
was
minimized with respect to two parameters, the  proton and neutron
harmonic-oscillator lengths. In the THO analysis, the energy minimization
involved the six parameters that define the LST functions for neutrons
$\{a_{n},b_{n},R_{n}\}$ and protons $\{a_{p},b_{p},R_{p}\}$. The calculations
were carried out for three
spherical doubly-magic nuclei, $^{16}$O, $^{40}$Ca,
and $^{208}$Pb.

Table \ref{tab01} contains the results for ground-state binding energies and
proton and neutron rms radii. As expected, the  binding energies based on THO
Slater determinants are lower than those based on the HO.  The HF binding
energies are of course lower than those based on either the HO or THO Slater
determinants because of the self-consistent nature of HF calculations (the HF
state is the optimal Slater determinant for a given Hamiltonian). In all
cases
considered, the THO binding energies are within 1.1\% of the HF results,
while
the HO results deviate up to 5\% for heavy nuclei. For neutron and proton rms
radii, the THO results agree very well with the self-consistent HF values,
whereas the HO results systematically overestimate them.

Table~\ref{tab02} compares the neutron single-particle energies that resulted
from the restricted HO and THO variational calculations with those from the
self-consistent HF calculations. The restricted calculations give a rather
good
approximation to the self-consistent single-particle energies. In general,
the
best agreement is obtained for the high-$\ell$ states where the tail of the
wave
function plays little role. The largest deviations are seen for $s$ and $p$
states.

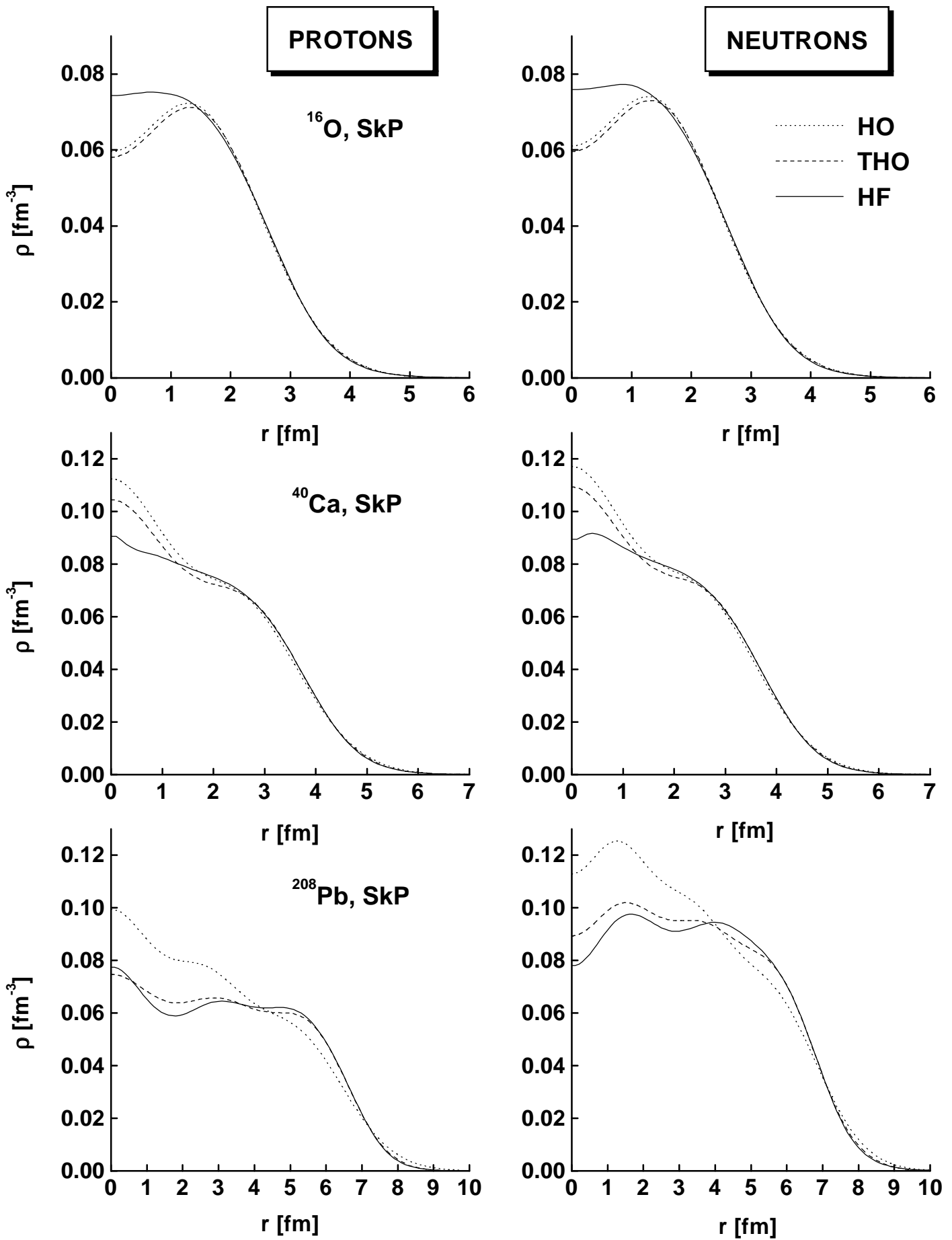
\begin{figure}
\caption{Local one body densities for $^{16}$O, $^{40}$Ca, and $^{208}$Pb
calculated in the self-consistent HF approach
(solid line) compared with
those obtained in restricted variational calculations using as trial wave
functions Slater determinants built from HO (dotted line) and THO (dashed
line)
single-particle orbitals. All calculations utilized the same SkP effective
interation. } \label{figHF}
\end{figure}

Finally, the neutron and proton density distributions resulting from the
various
calculations are compared in Fig.~\ref{figHF}. There is excellent agreement
between the THO and self-consistent HF densities in the surface region.  In
contrast, the incorrect asymptotic behavior of HO wave functions leads to
significant deviations \cite{[Ben89a]}. The THO densities differ most
significantly from the HF results at small values of $r$. However, since
the densities are weighted by $r^2$ in calculations of expectation values,
the main contribution to global nuclear characteristics such as energies and
rms
radii comes from the surface region.

\section{Summary and Conclusions}\label{conclusions}

In this paper, we have explored a new class of single-particle basis states
obtained by a local-scale point transformation (LST) of harmonic oscillator
states. We focussed special attention on the asymptotic properties of these
states (called THO states), to see whether they might be useful in the
description of weakly-bound nuclear systems, including those with a halo
structure.

Following a comprehensive summary of the LST formalism and its use in
building the THO basis, we discussed two applications of this new basis. The
first
concerned the description of sub-threshold (halo) states. We showed that the
THO
basis is greatly superior to the usual HO basis in reproducing the properties
of such weakly-bound states. We then discussed the use of this basis in
restricted HF calculations. Once again, the clear superiority of the THO
basis
to the ordinary HO basis was demonstrated, even for normal well-bound nuclei.
Most importantly, the optimal THO basis that emerged from the restricted HF
calculations provided an excellent reproduction of nuclear surface
properties.

The analysis presented in this paper should be viewed as a starting point
for future investigations. One potentially interesting application of the THO
basis is in the context of nuclear shell-model studies. The fact that the THO
basis expansion technique and the THO variational procedure accurately
reproduce
nuclear properties in the surface region suggests that the new basis can be
very
useful when studying those nuclear properties that depend on the asymptotic
behavior of the radial form-factor, and also for microscopic calculations of
the
effective interactions for weakly bound systems \cite{[Kuo97]}. Some of the
simplicity inherent in the use of HO wave functions in such studies will be
lost,
however. Calculations of G-matrix elements, for example, benefit greatly
from the
ease of transforming harmonic oscillator product wave functions into
relative and
center-of-mass coordinates, and this would be lost in the THO basis. On the
other
hand, it should be still much easier to carry out calculations in this basis
than
in a basis generated by numerical solution of the single-particle
Schr\"odinger
equation.

Another interesting avenue for future exploration concerns the use of the THO
basis in HFB or RHB calculations of weakly-bound nuclei where other
traditional methods (e.g., HF+BCS) cannot be used \cite{[Dob84],[Dob96]}. The
fact that with a rather modest number of basis THO states one can reproduce
the
properties of spatially extended states suggests that the traditional method
of
solving the HFB equations, based on basis expansion, can be revitalized by
using the THO basis. This way of solving the HFB problem (both in the
spherical
and deformed cases) can be an interesting alternative to algorithms based on
coordinate-space methods \cite{[Ter96],[Rei97],[Poe97]}.

Finally, the THO states that arise variationally can be used as
approximations to
the self-consistent canonical states of HFB. Recently, the variational method
based on the energy density formalism was generalized to the HFB case
\cite{[Sto98]}. In that work, the authors employed the basis proposed by
Ginocchio \cite{[Gin85]} in the description of semi-magic nuclei. Despite
the successes achieved in those calculations and in earlier
restricted HF calculations for doubly-magic nuclei \cite{[Sto97]}, there
are several features of the Ginocchio potential that make its more detailed
use problematical. First, since the centrifugal term in the Ginocchio
potential
is not treated properly, the resulting  wave functions with $\ell$$>$0 do not
have proper asymptotic behavior. Second, since the Ginocchio potential is
finite,
its bound eigenvectors do not form a closed set, and the use of the continuum
wave functions is necessary (see recent Ref. \cite{[Ben98a]}). In this
context,
the discrete THO basis would appear to be more useful.

\acknowledgments
Useful discussions with Jacek Dobaczewski are gratefully
acknowledged.
This research was supported in part by the Bulgarian National Foundation for
Scientific Research under contract No.~$\Phi $-527, the U.S. Department of
Energy under Contract Nos. DE-FG02-96ER40963 (University of Tennessee),
DE-FG05-87ER40361 (Joint Institute for Heavy Ion Research), DE-AC05-96OR22464
with Lockheed Martin Energy Research Corp. (Oak Ridge National Laboratory),
and
by the U. S. National Science Foundation under Grant Nos. PHY-9600445 and
INT-9722810.

\newpage

\begin{table}
\caption{Total binding energies and neutron and proton rms radii
obtained using HO and THO Slater determinants and from self-consistent HF
calculations for the spherical doubly-magic nuclei
 $^{16}$O, $^{40}$Ca, and $^{208}$Pb.}
\begin{tabular}{cccccccccccc}
& \multicolumn{3}{c}{$^{16}$O} &  & \multicolumn{3}{c}{$^{40}$Ca} &  &
\multicolumn{3}{c}{$^{208}$Pb} \\
& HO & THO & HF &  & HO & THO & HF &  & HO & THO & HF \\ \hline
$E$ (MeV) & -125.9 & -126.2 & -127.6 &  & -339.1 &
-340.9 & -343.3 &  & -1565.5 & -1621.8 & -1636.5 \\
$\langle r_p\rangle_{\rm rms}$~(fm) & 2.73 & 2.73 & 2.73 &  & 3.46 & 3.45 &
3.45 &  & 5.49 &
5.47 & 5.47 \\
$\langle r_n\rangle_{\rm rms}$~(fm) & 2.71 & 2.71 & 2.70 &  & 3.42 & 3.40 &
3.40
&  & 5.64 & 5.61 & 5.61
\end{tabular}
\label{tab01}
\end{table}

\begin{table}
\caption{Spherical single-neutron energies (in MeV)
obtained using HO and THO Slater determinants and from self-consistent HF
calculations for the spherical doubly-magic nuclei
$^{16}$O, $^{40}$Ca, and $^{208}$Pb.}
\begin{tabular}{cccccccccccc}
& \multicolumn{3}{c}{$^{16}$O} &  & \multicolumn{3}{c}{$^{40}$Ca} &  &
\multicolumn{3}{c}{$^{208}$Pb} \\
$nlj$ & HO & THO & HF &  & HO & THO & HF &  & HO & THO & HF \\ \hline
1$s_{1/2}$ & -29.0 & -29.2 & -29.0 &  & -36.4 & -37.1 & -37.4  &  & -37.1 &
-39.0 & -40.0 \\
1$p_{3/2}$ & -17.9 & -17.9 & -18.3 &  & -29.0 & -29.1 & -28.9 &  & -34.8 &
-35.6
& -36.4 \\
1$p_{1/2}$ & -13.5 & -13.6 & -13.6 &  & -25.5 & -25.5 & -25.8 &  & -33.3 &
-35.1
&  -35.8 \\
1$d_{5/2}$ &  &  &  &  & -19.6 & -19.6 & -19.7 &  & -31.3 & -31.6 &  -31.9 \\
2$s_{1/2}$ &  &  &  &  & -14.9 & -15.5 & -16.4 &  & -28.7 & -29.8 &  -29.2 \\
1$d_{3/2}$ &  &  &  &  & -14.5 & -14.5 & -14.4 &  & -29.0 & -30.4 &  -30.6 \\
1$f_{7/2}$  &  &  &  &  &  &  &  &  & -26.7 & -26.7 &  -26.9 \\
2$p_{3/2}$ &  &  &  &  &  &  &  &  & -23.2 & -23.8 &  -23.3 \\
1$f_{5/2}$ &  &  &  &  &  &  &  &  & -23.5 & -24.5 & -24.5 \\
2$p_{1/2}$ &  &  &  &  &  &  &  &  & -21.7 & -22.7 & -22.4 \\
1$g_{9/2}$  &  &  &  &  &  &  &  &  & -21.1 & -21.1 &  -21.3 \\
2$d_{5/2}$ &  &  &  &  &  &  &  &  & -17.1 & -17.5 &  -17.2 \\
1$g_{7/2}$ &  &  &  &  &  &  &  &  & -17.0 & -17.6 &  -17.6 \\
3$s_{1/2}$  &  &  &  &  &  &  &  &  & -15.0 & -15.8 &  -15.5 \\
2$d_{3/2}$ &  &  &  &  &  &  &  &  & -14.6 & -15.5 &  -15.6 \\
1$h_{11/2}$  &  &  &  &  &  &  &  &  & -14.8 & -15.0 &  -15.3 \\
2$f_{7/2}$ &  &  &  &  &  &  &  &  & -11.0 & -11.3 &  -10.9 \\
1$h_{9/2}$  &  &  &  &  &  &  &  &  & -9.7 & -10.0 & -10.1 \\
1$i_{13/2}$ &  &  &  &  &  &  &  &  & -8.0 & -8.5 &  -8.8  \\
3$p_{3/2}$  &  &  &  &  &  &  &  &  & -9.0 & -9.5 & -8.9 \\
2$f_{5/2}$  &  &  &  &  &  &  &  &  & -7.8 & -8.5 & -8.6 \\
3$p_{1/2}$ &  &  &  &  &  &  &  &  & -7.7 & -8.4 &  -8.1
\end{tabular}
\label{tab02}
\end{table}

\end{document}